\documentclass{revtex4-1}
\usepackage[latin9]{inputenc}
\usepackage{color,graphicx,epsfig}
\usepackage{ifpdf}
\usepackage{amsmath}
\usepackage{bm}
\usepackage{color}
\usepackage[english]{babel}
\usepackage{graphicx}%
\usepackage{amsfonts}%
\usepackage{amssymb}
\usepackage{braket}
\usepackage{hyperref}

\definecolor{nicered}{rgb}{0.7,0.1,0.1}
\definecolor{nicegreen}{rgb}{0.1,0.5,0.1}
\hypersetup{colorlinks,citecolor= nicegreen,linkcolor= nicered}

\makeatletter

\DeclareFontEncoding{LGR}{}{}
\DeclareTextSymbol{\~}{LGR}{126}
\newcommand{\lyxmathsym}[1]{\ifmmode\begingroup\def\b@ld{bold}
  \text{\ifx\math@version\b@ld\bfseries\fi#1}\endgroup\else#1\fi}

\@ifundefined{textcolor}{}
{%
 \definecolor{BLACK}{gray}{0}
 \definecolor{WHITE}{gray}{1}
 \definecolor{RED}{rgb}{1,0,0}
 \definecolor{GREEN}{rgb}{0,1,0}
 \definecolor{BLUE}{rgb}{0,0,1}
 \definecolor{CYAN}{cmyk}{1,0,0,0}
 \definecolor{MAGENTA}{cmyk}{0,1,0,0}
 \definecolor{YELLOW}{cmyk}{0,0,1,0}
}

\makeatother

\begin{document}


\title{Chiral loops in the isospin violating decays of $D_{s1}(2460)^+$ and $D_{s0}^*(2317)^+$ }
\author{ Svjetlana Fajfer}
\email[Electronic address:]{ svjetlana.fajfer@ijs.si}
\affiliation{Department of Physics,
  University of Ljubljana, Jadranska 19, 1000 Ljubljana, Slovenia}
\affiliation{J. Stefan Institute, Jamova 39, P. O. Box 3000, 1001
  Ljubljana, Slovenia}

\author{ A. Prapotnik Brdnik}
\email{ anita.prapotnik@um.si}
\affiliation{Faculty of Civil Engineering, Transportation Engineering and Architecture, University of Maribor, Smetanova ulica 17,
2000 Maribor, Slovenia}

\begin{abstract}
Positive parity meson states $D^*_{s0}(2317)^+$ and $D_{s1}(2460)^+$ have masses slightly below the $DK$ threshold. Both states can strongly decay only into isospin violating decays $D_{s1}(2460)^+ \to D_s^+ \pi \pi$, $D_{s1}(2460)^+ \to D^{*+}_s \pi^0$ and $D^*_{s0}(2317)^+ \to D_s^+ \pi^0$.
The $\pi$ states have rather small energies, which makes these decays   appropriate to study within  heavy meson chiral perturbation theory and calculate loop contributions. 
The $D_{s1}(2460)^+ \to D_s^+ \pi \pi$  decays occur only at  the loop level. Amplitude is a result of chiral loop contributions, which then have to be  finite. However, 
in the case of $D_{s1}(2460)^+ \to D_s^{*+} \pi^0$ and $D^*_{s0}(2317)^+ \to D_s^+ \pi^0$ decays, there is a tree-level contribution. We find that chiral loop contributions might  be important in both cases. The calculated amplitudes are  sensitive on the coupling constant describing the interaction of positive and negative  parity heavy meson multiplets with the light pseudoscalars. 
The counterterms contributions are also present in the amplitudes $D_{s1}(2460)^+ \to D_s^{*+} \pi^0$ and $D^*_{s0}(2317)^+ \to D_s^+ \pi^0$. We explore an experimentally known  ratio of the decay widths for these two decay modes to estimate the  size of counterterms  contributions. We  determine decay widths for both decay modes to be $\Gamma (D_{s1}(2460)^+ \to D_s^+ \pi^+ \pi^-) \simeq 0.25\,$keV and  $\Gamma (D_{s1}(2460)^+ \to D_s^+ \pi^0 \pi^0 ) \simeq 0.15\,$keV.
\end{abstract}
\maketitle
\section{Introduction}

It has been more than a decade since $D^*_{s0}(2317)^+$ and $D_{s1}(2460)^+$  mesons  were observed  by the BABAR and CLEO collaborations,  respectively. Their existence  was confirmed by several
experiments \cite{Aubert:2003fg}-\cite{Aubert:2003pe}. The measurement of  $D^*_{s0}(2317)^+$ and $D_{s1}(2460)^+$ quantum numbers, as well as decay widths and decay rates, has continued since that time \cite{Aubert:2003fg}-\cite{Abe:2003jk}. The experimental data support the interpretation of  the $D^{*}_{s0}(2317)^+$ meson as  a positive parity scalar (${\rm J^P}=0^+$),  while $D_{s1}(2460)^+$ appears to be positive parity axial vector (${\rm J^P}=1^+$). 
Both states behave as $c\bar s$ systems, although  their masses  turned out to be 100 MeV smaller than expected by calculation based on the quark models (for a review see \cite{Colangelo:2004vu}).
 Many proposals have suggested that these states are tetraquarks or  {\it DK} molecules  \cite{Colangelo:2004vu}-\cite{Navarra:2015iea}.
 There is also a suggestion that these states can  be a result of the  mixing of $\bar s c$ and $\bar s c \bar q q$ states (see  e.g. \cite{Browder:2003fk}). Recently,  
 lattice studies indicated that  $D^*_{s0}(2317)^+$  is a combined  state of $\bar s c$  and $DK$ molecules \cite{Liu:2012zya,Mohler:2013rwa,Lang:2014yfa,Mohler:2011ke}. However, multiple lattice volumes will be needed \cite{Lang:2014yfa} to resolve the structure  of $D^*_{s0}(2317)^+$. On the other hand, the total decay widths of $D^*_{s0}(2317)^+$ and $D_{s1}(2460)^+$  are still  unmeasured. The upper bound on the total decay width of the $D^*_{s0}(2317)^+$ meson is $3.5\,$MeV at 95\% confidence level, while the upper bound on the total decay width of the $D_{s1}(2460)^+$ meson is $3.8\,$MeV at 95\% confidence level \cite{Aubert:2006bk}. Nevertheless, some branching rations for $D_{s1}(2460)^+$ decays  were determined. The masses of  $D^*_{s0}(2317)^+$ and $D_{s1}(2460)^+$ mesons are slightly below the threshold for the decay into a {\it D} and a {\it K} meson. Therefore, only strong isospin violating decays of $D^*_{s0}(2317)^+$ and $D_{s1}(2460)^+$   are kinematically allowed, as well as radiative decays of $D^*_{s0}(2317)^+$ and $D_{s1}(2460)^+$. 
The decay  $D^*_{s0}(2317) \to D_s^+\pi^0$ was observed by the BABAR collaboration \cite{Aubert:2003fg}, but branching ratio or partial decay width was not determined. However, the branching ratio of $D_{s1}(2460)^+ \to D_s^{*+} \pi^0$ was found to be $(48\pm 11) \% \,$  \cite{Aubert:2006bk,PDG}. Thr BABAR Collaboration  also observed $D_{s1}(2460)^+ \to D_s^+ \pi^+ \pi^-$  \cite{Aubert:2006nm}, with the branching ratio of $(4.3\pm 1.3)\,$\% \cite{PDG}. There is no measurement of the branching fraction for  the decay mode $D_{s1}(2460)^+ \to D_s^+ \pi^0 \pi^0$ yet.  
In the case that  $D^*_{s0}(2317)^+$ and $D_{s1}(2460)^+$ are only   $c\bar s$ mesons, with quantum numbers   $0^+$ and $1^+$, the decay of $D^*_{s0}(2317)^+$ into $D_s^+ \pi^0$ is possible, while the decays of 
$D^*_{s0}(2317)^+$ into $D_s^{*+} \pi^0$ and $D_s^+ \pi\pi$ cannot occur. Similarly, the decay of 
$D_{s1}(2460)^+$ into $D_s^+\pi^0$ is not possible, while the decays of $D_{s1}(2460)^+$ into $D_s^{*+} \pi^0$ and $D_s^+ \pi\pi$ are allowed. 
Therefore, it seems that existing experimental results favor the treatment of both states  $D^*_{s0}(2317)^+$ and $D_{s1}(2460)^+$ as a $c \bar s$ states with the positive parity.  

The $D_{s1}(2460)^+ \to D_s^{*+} \pi^0$ and $D^*_{s0}(2317)^+ \to D_s^+ \pi^0$ decays were considered within variety of approaches, \cite{Godfrey:2003kg}-\cite{Wei:2005ag}.   
The quark models were mostly  exploited  for the decays $D^*_{s0}(2317)^+ \to D_s^+ \eta \to D_s^+ \pi^0$ and $D_{s1}(2460)^+ \to D_s^{*+} \eta \to D_s^{*+} \pi^0$, assuming isospin symmetry violation via $\eta-\pi$ mixing. In Ref. \cite{Faessler:2007gv} states $D_{s1}(2460)^+$ and 
$D^*_{s0}(2317)^+$ are treated in the $DK$- molecule picture. Most of these studies predict the partial decay widths of  $D_{s1}(2460)^+ \to D_s^{*+} \pi^0$ and $D^*_{s0}(2317)^+ \to D_s^+ \pi^0$ decay modes to be in the range  $(10-30)\,$keV, although higher values (about $100\,$keV) were suggested  in Ref. \cite{Azimov:2004xk} too. 

The three-body decays $D_{s1}(2460)^+ \to D_s^+ \pi^+ \pi^-$  and $D_{s1}(2460)^+ \to D_s^+ \pi^0 \pi^0$ were studied in Refs. \cite{Bardeen:2003kt,Lu:2006ry}. The authors of  \cite{Bardeen:2003kt} assumed that the decay occurs trough the intermediate $\tilde \sigma$ fields, which then convert to pions through the $\bar ss$  component of $\tilde \sigma$ fields mixing with the $\bar uu$ and $\bar dd$ components. Two relevant $\sigma$ states ($\sigma_0$ and $\sigma_8$) were considered with masses set to $1\,$GeV and $1.5\,$GeV. 
In Ref.  \cite{Bardeen:2003kt}   high sensitivity of the amplitude on  the mass of the lighter sigma meson state was found out and the variation of its mass in the range  $0.8\,$GeV and $1.2\,$GeV can change the predicted result for an order of magnitude. 
In \cite{Lu:2006ry}, the $\sigma$ states were replaced by the scalar   $f_0(980)$ state. 
However, the decay width was  estimated to be  $\Gamma(D_{s1}(2460)^+ \to D_s^+ \pi^+ \pi^-)<25\,$keV.

In this paper, we determine chiral loop contributions  to the isospin violating decay amplitudes of $D^{*}_{s0}(2317)^+$ and $D_{s1}(2460)^+$ mesons. 
For two-body decays, there is a tree-level  contribution to  decay  amplitude  arising from the $\eta$-$\pi$ mixing. However,  chiral loops even in these cases, might give significant contributions. This was indicated already in Ref. \cite{Cleven:2014oka}, where some of the loop contributions to the decay  amplitude   were determined.  In our analysis,  we rely on the heavy meson chiral perturbation theory (HM$\chi$PT). Within HM$\chi$PT the  $D^{*}_{s0}(2317)^+$ and $D_{s1}(2460)^+$ states have  quantum numbers of  $c \bar  s$.
The  use of HM$\chi$PT in both decay modes is fully justified by the fact that the pions in the final set have rather small energies.  The loop contributions  within this framework arise from the light pseudoscalar meson exchanges. The  light resonances, as light vector mesons  ($\rho$, $K^*$) in the amplitudes  at tree level, give the contributions of the same order in  the chiral counting \cite{Weinberg:1991um} as light pseudoscalar  meson loops \cite{Ecker:1988te}. 
In comparison with the approach  of \cite{Cleven:2014oka}, we find  that there are additional Feynman diagrams leading to the relevant contribution to the two body decay amplitudes. The three body decay amplitude within this framework arises from chiral loops. The energy release in both two-body and three-body decays is very  small. Both negative and positive parity intermediate  $D$ states are taken into account  within this framework \cite{Stewart:1998ke,Fajfer:2006hi,Becirevic:2004uv}. 

The heavy meson Lagrangian formalism will be introduced in Sec. II. In Sec. III, the analysis of the $D_{s1}(2460)^+ \to D_s^+ \pi \pi$ decay channels will be presented, while
in Sec. IV we will calculate the decay width of the two-body   $D_{s1}(2460)^+ \to D_s^{*+} \pi^0$ and $D^*_{s0}(2317)^+ \to D_s^+ \pi^0$ decay modes while a short conclusion will be given in Sec. V.

\section{Framework}
\label{framework}

The framework of heavy meson chiral perturbation theory combines the heavy quark effective theory with the
chiral perturbation theory \cite{Burdman:1992gh,Wise:1992hn}. Heavy quark effective theory is used to describe mesons composed of one heavy quark (c or b) and one light quark (u, d or s) \cite{Burdman:1992gh,Wise:1992hn}. In such mesons, the heavy quark moves almost on shell with the velocity $v^\mu$ and  the momenta of the heavy meson can be written as $p^\mu=m v^\mu+k^\mu$, where $m$ is a heavy meson mass and $k^\mu$ is of the order of $\Lambda_{QCD}$ and much smaller then $m v^\mu$. In a limit, when the mass of the heavy quark becomes infinite, pseudoscalar and vector meson states become degenerate,
as well as scalar and axial vector meson states. The negative parity states are described by the field $H$, while the positive parity states are described by the field $S$:
\begin{equation}
H=\frac{1}{2}(1+v \cdot \gamma)[P^*_\mu \gamma^\mu-P\gamma_5]\,, \qquad S=\frac{1}{2}(1+v \cdot \gamma)[D^*_\mu \gamma^\mu\gamma_5-D]\,,
\end{equation}
where $P^*_\mu$ and $P$ annihilate the vector and pseudoscalar mesons, respectively, while  $D^*_\mu$ and $D$ annihilate the axial-vector and scalar mesons, respectively.

Within chiral perturbation theory, the light pseudoscalar mesons are accommodated into the octet  $\Sigma=\xi^2=e^{(2i\Pi/f)}$ with
\begin{equation}
\Pi=
\left(
\begin{matrix}
\pi^0/\sqrt{2}+\eta_8/\sqrt{6} & \pi^+ & K^+ \\
\pi^- & -\pi^0/\sqrt{2}+\eta_8/\sqrt{6} & K^0 \\
K^- & \bar K^0 & -2\eta_8/\sqrt{6}
\end{matrix} \right)
\label{eq-pi}
\end{equation}
and $f\sim 120\,$MeV at one loop \cite{f}.
The leading order of the HM$\chi$PT Lagrangian, that describes the interaction of heavy and light mesons, can be written as
$${\cal L}=-
Tr[\bar H_a (iv\cdot {\cal D}_{ab}-\delta_{ab}\Delta_H)H_b]+gTr[\bar H_bH_a\gamma \cdot {\cal A}_{ab}\gamma_5]$$
\begin{equation}
+Tr[\bar S_a (iv\cdot {\cal D}_{ab})-\delta_{ab}\Delta_S)S_b]+\tilde g Tr[\bar S_b S_a\gamma \cdot {\cal A}_{ab}\gamma_5]+hTr[\bar H_bS_a\gamma \cdot{\cal A}_{ab}\gamma_5]\,,
\label{eq-lagrange}
\end{equation}
where ${\cal D}_{ab}^\mu=\delta_{ab}\partial^\mu-{\cal V}_{ab}^\mu$ is a heavy meson covariant derivative, ${\cal V}_\mu =1/2(\xi^\dagger \partial_\mu \xi +\xi \partial_\mu \xi^\dagger)$ is the light meson vector current, and  ${\cal A}_\mu =i/2(\xi^\dagger \partial_\mu \xi -\xi \partial_\mu \xi^\dagger)$ is the light meson axial current. A trace is taken over spin matrices and repeated light quark flavor indices.  
All terms in (\ref{eq-lagrange}) are of the order ${\cal O}(p)$  in  the chiral power counting (see, e.g.,\cite{Fajfer:2006hi}). As in \cite{Fajfer:2006hi} we assign for $\Delta_{SH} = \Delta_S -\Delta_H  \sim {\cal O}(p)$ in order 
 to maintain a well-behaved  chiral expansion.  Light mesons are described by the Lagrangian \cite{Burdman:1992gh,Wise:1992hn},  which is of the order ${\cal O}(p^2)$ in the chiral expansion:
\begin{equation}
{\cal L}_0=\frac{f^2}{8}Tr[\partial_\mu \Sigma \partial^\mu \Sigma^\dagger]+\frac{f^2\lambda_0}{4}Tr[m_q^\xi \Sigma+\Sigma m_q^\xi]\,,
\label{eq-light}
\end{equation}
with $\lambda_0={m_\pi^2}/{(m_u+m_d)}={(m^2_{K^+}-m^2_{K_0})}/{(m_u+m_d)}={(m^2_K-m^2_\pi/2)}/{m_s}\, $.
The above Lagrangians lead to Feynman rules, as  given in \cite{Fajfer:2006hi}. 
The scalar (pseudoscalar) and vector (axial-vector) heavy meson propagators can be written in the  forms
\begin{equation}
\frac{i}{2(k\cdot v-\Delta_i)} \qquad {\rm and} \qquad \frac{-i(g^{\mu\nu}-v^\mu v^\nu)}{2(k\cdot v-\Delta_i)}
\end{equation} 
respectively, where $\Delta_i$ in the propagator represents the residual mass of the corresponding field. Residual masses are responsible for mass splitting of heavy meson states. The difference $\Delta_S-\Delta_H$ splits the masses of positive and negative parity states. In addition, we also have to take into account mass splitting between $D_s$ and $D$ states as well as mass splitting between vector (axial-vector) and pseudoscalar (scalar) fields. These splittings arise due to the heavy meson Lagrangian correction of the order ${\cal O}(m_q)$ ($m_q$ stands for the mass of light quarks). To account for all of the above mass splittings, we will follow the approach of 
\cite{Stewart:1998ke,Fajfer:2006hi} and set the values of $\Delta_i$ to the experimentally measured mass differences between $D$ meson states. We use the mass of the initial particle as a reference value, so all mass differences are defined as mass differences between the relevant $D$ meson and initial state \cite{Stewart:1998ke}. 

The coupling constants $g$, $h$, and $\tilde g$ were already discussed by several authors and determined by several methods: 
the QCD sum rules \cite{Colangelo:1997rp}-\cite{Colangelo:1994es}, the lattice  QCD \cite{Becirevic:2012pf}-\cite{Blossier:2014vea}, and the extraction from the experimental data \cite{Fajfer:2006hi,Stewart:1998ke,Anastassov:2001cw,Colangelo:2005gb}. We will use  recent results from  the lattice QCD: $g=0.54(3)(^{+2}_{-4})$ \cite{Becirevic:2012pf}, $\tilde g=-0.122(8)(6)$, and $h=-0.84(3)(2)$ \cite{Blossier:2014vea}. The values of $h$ and $\tilde g$ were determined for the $B$ meson sector, so $1/m_c$ corrections can make a slight difference in stated values.  Although HM$\chi$PT relies on the  expansion in the light pseudoscalar momentum and 
$1/m_c$ expansion, we  do not consider $1/m_c$  corrections for at least two reasons: first, the number of additional terms in the Lagrangian becomes huge and impossible to estimate and second, 
lattice studies  indicate that these contributions are rather small \cite{Dougall:2003hv}. 
The authors of \cite{Blossier:2014vea}
made an estimation of $h$ for  the $D$ meson sector and found that  $h\simeq 0.74(8)$,  which is still inside the error bars of $h=-0.84(3)(2)$. 

In order  to absorb divergences coming from  loop integrals, one needs to include  counterterms in the Lagrangian. Following \cite{Stewart:1998ke,Fajfer:2006hi}, they can be written as$$
{\cal L}_{ct}=
\lambda_1[\bar H_b\bar H_a(m_q^\xi)_{ba}]+\lambda^\prime_1[\bar H_a\bar H_a(m_q^\xi)_{bb}]
-\tilde{\lambda_1}[\bar S_b\bar S_a(m_q^\xi)_{ba}]-\tilde{\lambda^\prime_1}[\bar S_a\bar S_a(m_q^\xi)_{bb}]+
$$$$
\frac{h \kappa_1^\prime\lambda_0}{(4\pi f)^2}Tr[(\bar H S \gamma_\mu{\cal A}^\mu \gamma_5)_{ab}(m_q^\xi)_{ba}]+
\frac{h \kappa_3^\prime\lambda_0}{(4\pi f)^2}Tr[(\bar H S \gamma_\mu{\cal A}^\mu \gamma_5)_{aa}(m_q^\xi)_{bb}]+
$$$$
\frac{h\kappa_5^\prime\lambda_0}{(4\pi f)^2}Tr[\bar H_a S_a \gamma_\mu{\cal A}^\mu_{bc} \gamma_5(m_q^\xi)_{cb}]+
\frac{h\kappa_9^\prime\lambda_0}{(4\pi f)^2}Tr[\bar H_c S_a \gamma_\mu{\cal A}^\mu_{bc} \gamma_5(m_q^\xi)_{ab}]+
$$
\begin{equation}
\frac{\delta_2^\prime}{(4\pi f)^2}Tr[\bar H_a S_b i v\cdot {\cal D}_{bc}\gamma_\mu{\cal A}^\mu_{ca} \gamma_5]+
\frac{\delta_3^\prime}{(4\pi f)^2}Tr[\bar H_a S_b i \gamma_\mu\cdot {\cal D}_{bc}^\mu v \cdot {\cal A}^\mu_{ca} \gamma_5]+h.c.+\hdots\,,
\label{eq-counterterms}
\end{equation}
where $m^\xi=(\xi m_q \xi-\xi^\dagger m_q \xi^\dagger)/2$ and $D^\alpha_{bc}A^\beta_{ca}=\partial^\alpha A^\beta_{ba}+[v^\alpha A^\beta]_{ba}$. At the given scale, the finite part of $\kappa^\prime_3$ can be absorbed into the definition of $h$. Parameters $\lambda^\prime_1$ and $\tilde{\lambda^\prime_1}$ can be absorbed into the definition of heavy meson masses by phase redefinition of $H$ and $S$, while 
$\lambda_1$ and $\tilde{\lambda_1}$ split the masses of SU(3) flavor triplets of $H_a$ and $S_a$ \cite{Stewart:1998ke,Fajfer:2006hi}. Therefore, only  contributions proportional to $\kappa^\prime_1$, $\kappa^\prime_9$, $\kappa^\prime_5$, $\delta^\prime_2$, and  $\delta^\prime_3$ will be explicitly included in the amplitudes.

\section{The $D_{s1}(2460)^+ \to D_s^+ \pi^+ \pi^-$ and $D_{s1}(2460)^+ \to D_s^+ \pi^0 \pi^0$ decay modes}

\begin{figure}
\begin{center}
\includegraphics[scale=0.7]{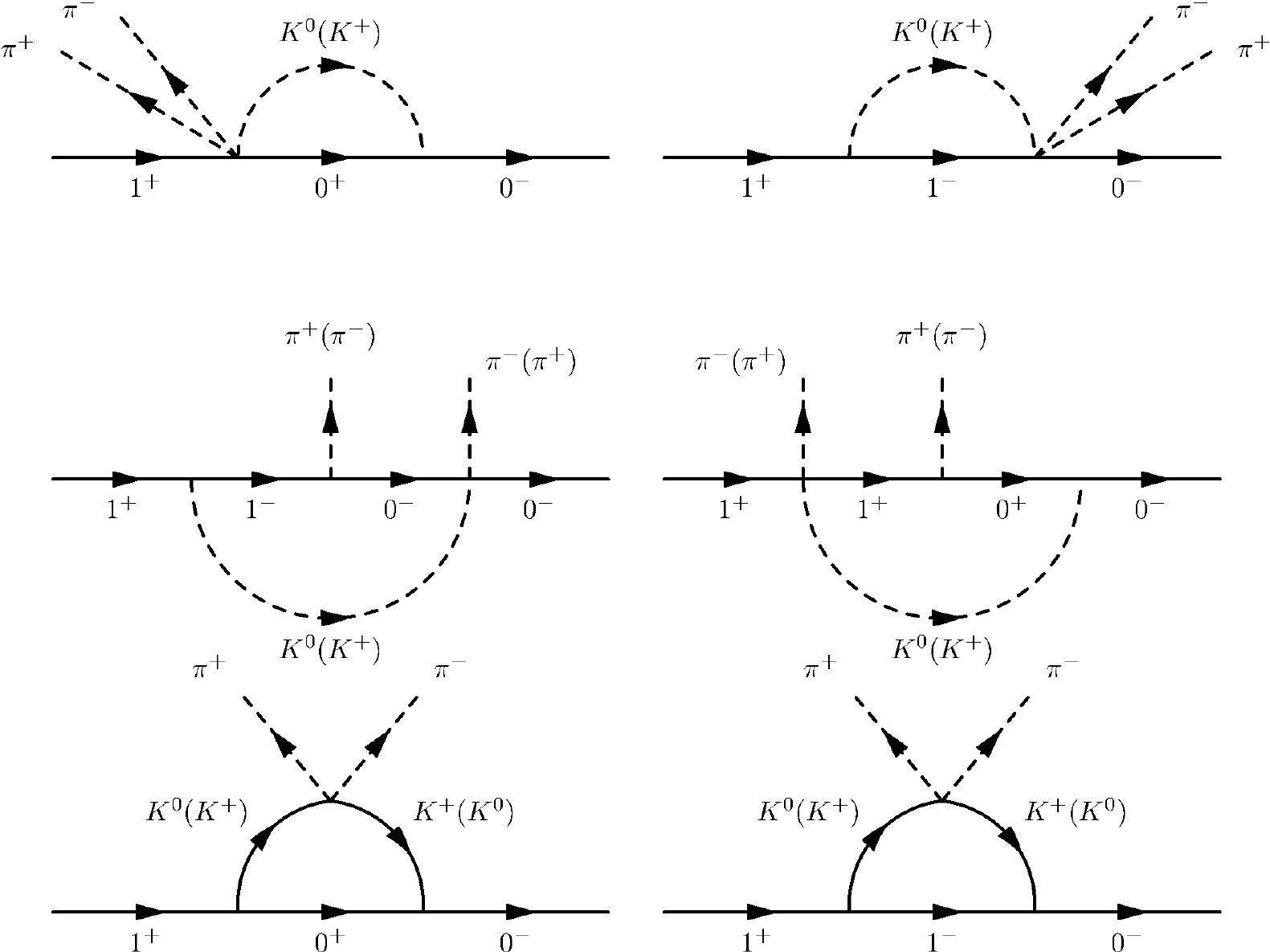}
\caption{Non-zero contributions  to $D_{s1}(2460)^+ \to D_s^+ \pi^+ \pi^-$ decay amplitude.} 
\label{grafi1}
\end{center}
\end{figure}

First, we will consider the chiral loop contributions to the $D_{s1}(2460)^+ \to D_s^+ \pi^+ \pi^-$ 
decay rate.
The decay width for the $D_{s1}(2460)^+ \to D_s^+ \pi^+ \pi^-$, averaged over the $D_{s1}(2460)^+$ polarizations, can be written as:
\begin{equation}
d\Gamma=\frac{1}{(2\pi)^3}\frac{1}{32 M_i^3}|{\cal M}|^2 dm^2_{12}dm^2_{23}\,,
\label{gama}
\end{equation}
where $M_i$ is the mass of initial particle  $D_{s1}(2460)^+$,  $dm^2_{12}=(p_++p_-)^2$, and $dm^2_{23}=(p_-+q)^2$. Here $p_-$, and $p_+$ denote the momenta of
$\pi^+$ and $\pi^-$ respectively, while $q$ is the momentum of $D_s^+$. 
The decay amplitude ${\cal M}$, in general, can be written in the form
$${\cal M}=\epsilon \cdot ({\cal A}_1p_+ +{\cal A}_2p_- + {\cal A}_3 q)\,.$$
Within heavy quark limit $P^\mu=M_i v^\mu$,  $q^\mu=M_fv^\mu$, and  $p_+^\mu+p_-^\mu=\Delta_M v^\mu$, where $\Delta_M=(M_i-M_f)$
and $M_f$ is the mass of  $D_s^+$ meson.
As $\epsilon \cdot v=0$, the amplitude simplifies to
$${\cal M}={\cal A}\, \epsilon \cdot (p_+-p_-)={\cal A}\, \epsilon \cdot \Delta p\,,$$
where ${\cal A}$ can be calculated from the diagrams presented in Fig. \ref{grafi1}.
Only diagrams giving a nonzero contribution are shown. 
Note, that there are no diagrams with the $\eta$ meson in the loop on Fig. \ref{grafi1}, as they all  give a vanishing contribution.

\begin{figure}
\begin{center}
\includegraphics[scale=0.7]{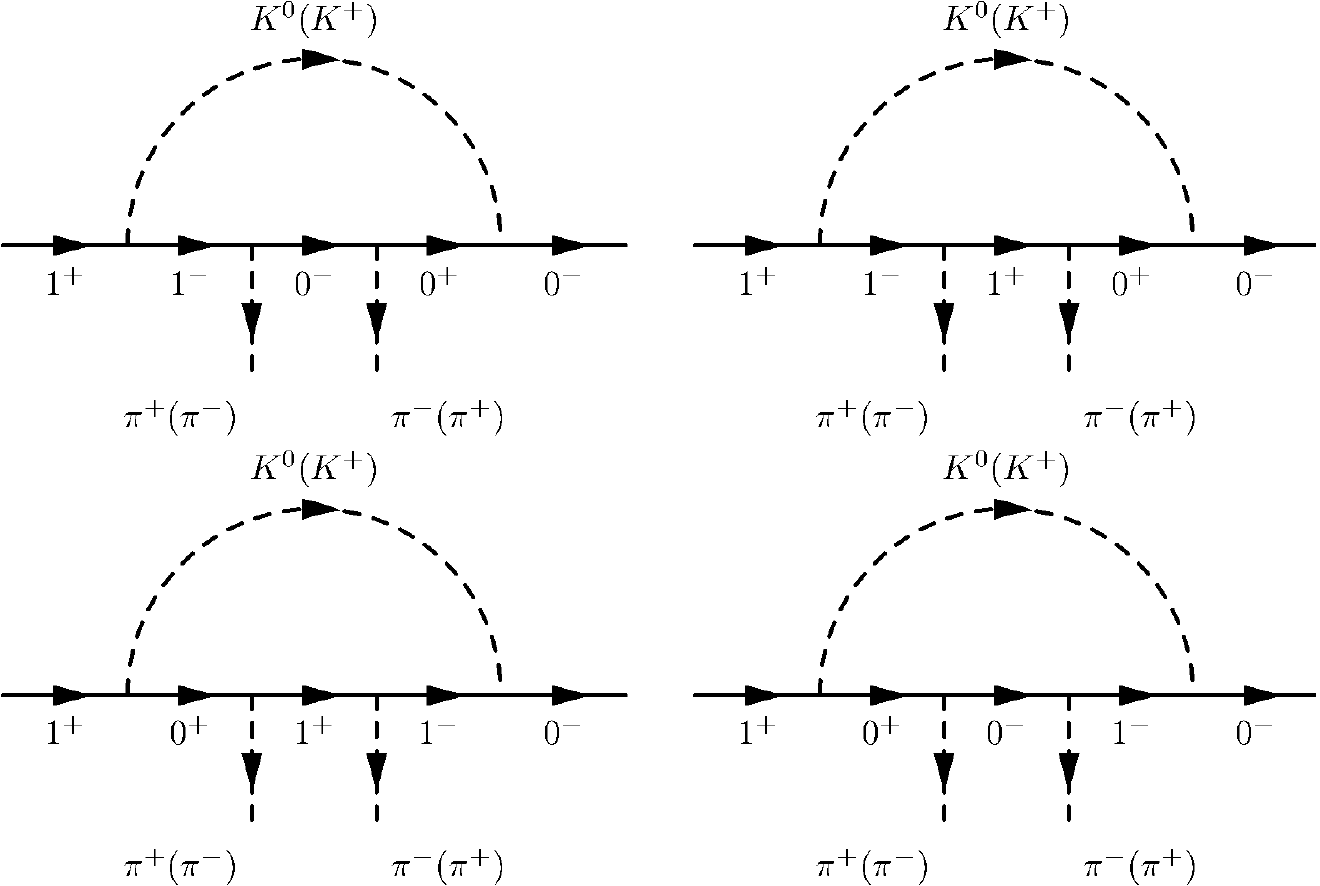}
\caption{Feynman diagrams contributing to $D_{s1}(2460)^+ \to D_s^+ \pi^+ \pi^-$  decay amplitude at higher order and which are not included in our calculations.}
\label{grafi2}
\end{center}
\end{figure}

The diagrams in Fig. \ref{grafi2} also give a nonzero contribution to the amplitude. 
Since they are next-to-leading order in HM$\chi$PT their contributions can be neglected.  Note, also, that by taking $P^\mu=M_iv^\mu$ and  $q^\mu=M_fv^\mu$, all scalar products of the momenta become independent of the phase space parameters: $p_+\cdot v=p_- \cdot v=\Delta M/2$ and $p_-\cdot p_+=\Delta M/2 -m^2$. Therefore, the spin averaging of the amplitude $\cal M$ is constant on the whole phase space region, implying that the calculation of the amplitude and the integration over the phase space in (\ref{gama}) can be done independently.

Now, we proceed to the calculation of the amplitude $\cal A$. Using  Feynman rules derived from  (\ref{eq-lagrange}), we obtain
\begin{equation}
{\cal A}=\frac{h\sqrt{M_iM_f}}{16\pi^2 f^4}\left(a_1+a_2+b_1+b_2+c_1+c_2\right)\,,
\end{equation}
where
\begin{equation}
a_1 = \frac{g}{2}\left(\bar B_1(\Delta_{P^*}, m_{K^0})-\bar B_1(\Delta_{P^*}, m_{K^+})\right)\,,
\end{equation}
\begin{equation}
a_2 =\frac{\tilde g}{2}\left(\bar B_1(\Delta_{D}+\Delta_M, m_{K^0})-\bar B_1(\Delta_{D}+\Delta_M, m_{K^+})\right)\,,
\end{equation}
\begin{equation}
b_1 = 2g\left(\left(\bar B_2^\prime(\Delta_{P^*},\Delta_P +\Delta_M/2, m_{K^0})
-\Delta_M/2\cdot \bar B_1^\prime(\Delta_{P^*},\Delta_P +\Delta_M/2, m_{K^0})\right)\right.$$$$\left.
-\left(\bar B_2^\prime(\Delta_{P^*},\Delta_P +\Delta_M/2, m_{K^+})
-\Delta_M/2\cdot \bar B_1^\prime(\Delta_{P^*},\Delta_P +\Delta_M/2, m_{K^+})\right)
\right)\,,
\end{equation}
\begin{equation}
b_2 = 2\tilde g\left(\left(\bar B_2^\prime(\Delta_{D^*}+\Delta_M/2,\Delta_D  +\Delta_M, m_{K^0})
+\Delta_M/2\cdot \bar B_1^\prime(\Delta_{D^*} +\Delta_M/2,\Delta_D  +\Delta_M, m_{K^0})\right) \right. $$$$
-\left.\left(\bar B_2^\prime(\Delta_{D^*}+\Delta_M/2,\Delta_D+\Delta_M, m_{K^+})
+\Delta_M/2\cdot \bar B_1^\prime(\Delta_{D^*},\Delta_D+\Delta_M, m_{K^+})
\right)\right)\,,
\end{equation}
\begin{equation}
c_1=-2g\left(\left(B_{00}(m_{K^0})+\Delta_{P^*}\bar C_{00}(\Delta_{P^*},m_{K^0})\right)-\left(B_{00}(m_{K^+})+\Delta_{P^*}\bar C_{00}(\Delta_{P^*},m_{K^+})\right)\right)$$$$
c_2=-2\tilde g\left(\left(B_{00}(m_{K^0})+\Delta_{D}\bar C_{00}(\Delta_{D},m_{K^0})\right)-\left(B_{00}(m_{K^+})+\Delta_{D}\bar C_{00}(\Delta_{D},m_{K^+})\right)\right)\,.
\end{equation}
Here,  $\bar B_1$, $\bar B_2$, $\bar B_{00}$ and $\bar C_{00}$ are the   Passarino-Veltman loop integrals defined in Appendix A. Note that these loop integrals, as well as the parts of the amplitude $a_i$, $b_i$, and $c_i$, are divergent in the $D\rightarrow 4$ limit. Nevertheless, by summing up all contributions,  divergences cancel out, so both $\cal A$ and $\cal M$ are finite. As $D_{s1}(2460)^+ \to D_s^+ \pi^+ \pi^-$ does not have any tree-level contributions from the heavy meson Lagrangian, this was expected. Another interesting feature is, that the amplitude $\cal M$ is nonzero due to the mass difference of the $K^+$ and $K^0$ mesons. Namely, if we put $m_{K^+}=m_{K^0}$, the amplitude would vanish.
Finally, the decay width coming from these amplitudes is
$$\Gamma(D_{s1}(2460)^+ \to D_s^+ \pi^+ \pi^-)=0.25(4)(7)(^{+2}_{-4})\,{\rm keV}\,.$$
The first error comes from the uncertainty in the coupling constant $h$, the second from uncertainty in the coupling constant $g$, and the last from the uncertainty in the mass of $D_{s1}(2460)^+$ meson. Uncertainties in the coupling constant $\tilde g$ and other meson masses are relatively small and therefore can be safely neglected. This result implies that the total decay width of $D_{s1}(2460)^+$
is found to be  between $2\,$keV and $13\,$keV.
 
The decay amplitude for  $D_{s1}(2460)^+ \to D_s^+ \pi^0 \pi^0$ can be easily found by replacing the $\pi^+ \pi^-$ state by $\pi^0\pi^0$ in the final state. Note that one has to include the factor 1/2 in the decay mode due to two identical bosons in the final state. 
This yields
$$\Gamma(D_{s1}(2460)^+ \to D_s^+ \pi^0 \pi^0)=0.15(4)\,{\rm keV}\,.$$

\section{The  $D_{s1}(2460)^+ \to D_s^{*+} \pi^0$ and $D^*_{s0}(2317)^+ \to D_s^+ \pi^0$ decay modes}

\begin{figure}
\begin{center}
\includegraphics[scale=1]{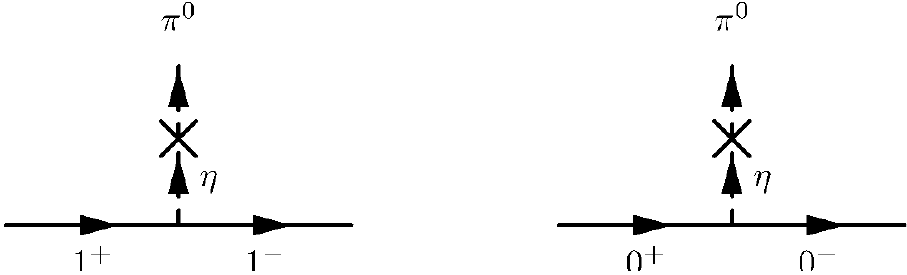}
\caption{Tree-level contribution to $D_{s1}(2460)^+ \to D_s^{*+} \pi^0$ and $D_{s0}^*(2317)^+ \to D_s^+ \pi^0$ decay modes.}
\label{tree}
\end{center}
\end{figure}

\begin{figure}
\begin{center}
\includegraphics[scale=0.7]{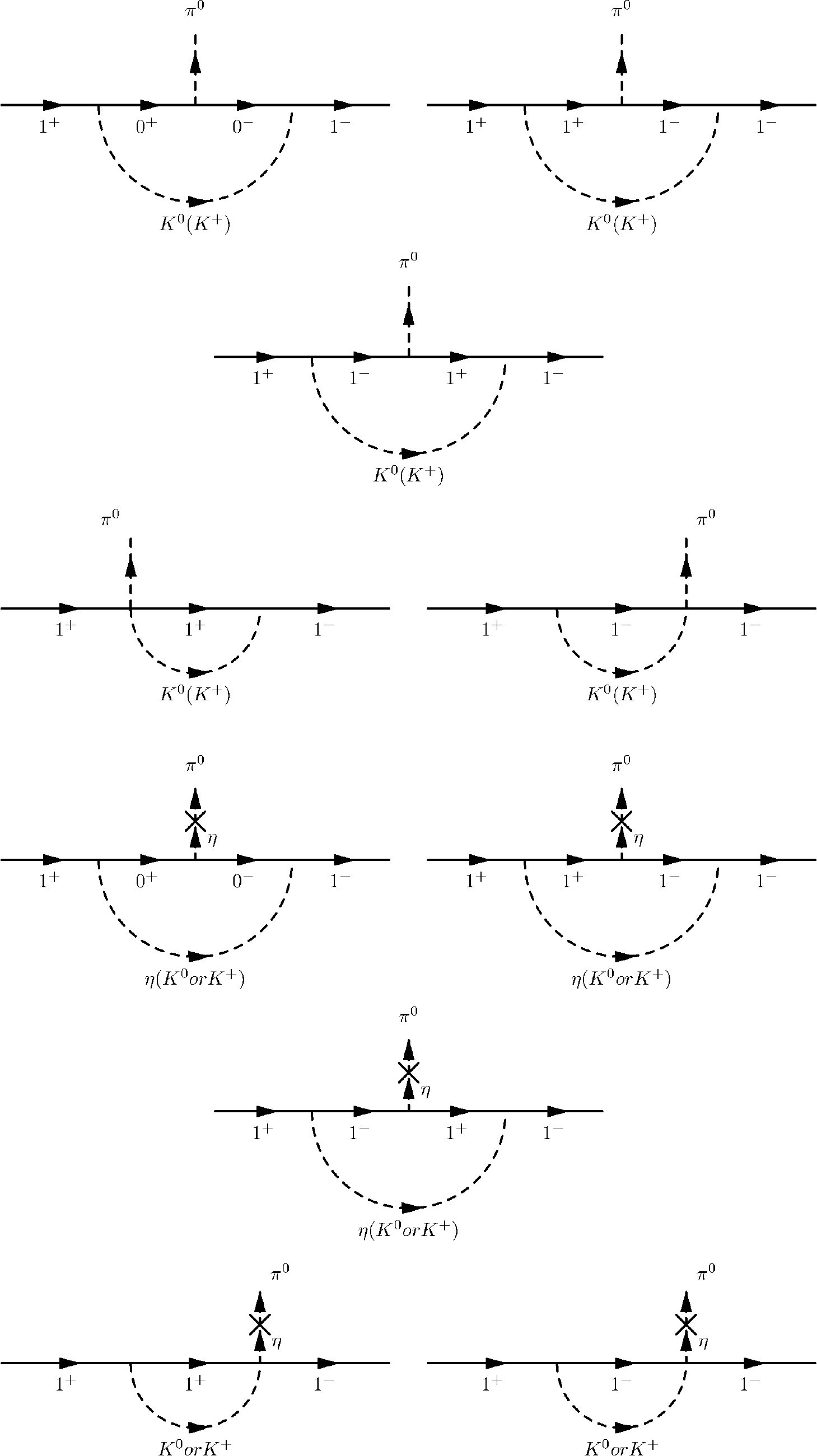}
\caption{Chiral corrections to the $D_{s1}(2460)^+ \to D_s^{*+} \pi^0$ decay mode.}
\label{chi1}
\end{center}
\end{figure}

The tree-level contribution to $D_{s1}(2460)^+ \to D_s^{*+} \pi^0$ and 
$D^*_{s0}(2317)^+ \to D_s^+ \pi^0$ decay amplitudes results from the $\eta-\pi$ mixing.
 In this scenario, the decays proceed  through the channels
$D_{s1}(2460)^+ \to D_s^{*+} \eta  \to D_s^{*+} \pi^0$ and 
$D^*_{s0}(2317)^+ \to D_s^+ \eta  \to D_s^+ \pi^0$ as presented in  Fig. \ref{tree}.
The $\eta-\pi^0$ mixing can be approached by  the mixing Lagrangian \cite{mixing1,mixing2},
\begin{equation}
{\cal L}_{\eta-\pi_0}=\frac{m_\pi^2 (m_u-m_d)}{\sqrt{3}(m_u+m_d)}\pi_0\eta\,,
\end{equation}
which comes from the second term in (\ref{eq-light}).
The decay width  for  $D_{s1}(2460)^+ \to D_s^{*+} \pi^0$  coming from this mixing tree-level amplitude is
\begin{equation}
\Gamma (D_{s1}(2460)^+ \to D_s^{*+} \pi^0) =\frac{h^2}{2\pi f^2}|k_\pi|E_\pi \delta^2_{mix}=16\, {\rm keV}\,,
\end{equation}
where $k_\pi$ and $E_\pi$ are the momenta and energy of the outgoing pion,   
 while mixing parameter $\delta_{mix}$ is defined as  in \cite{mixing1,mixing2}
\begin{equation}
\delta_{mix}=\frac{1}{2\sqrt{2}}\frac{m_u-m_d}{m_s-(m_u+m_d)/2}\,.
\end{equation}
 Note that decay widths for  $D_{s1}(2460)^+ \to D_s^{*+} \pi^0$ and $D^*_{s0}(2317)^+ \to D_s^+ \pi^0$ decay modes differ only for a small difference in $D_s^+$  masses. 

\begin{figure}
\begin{center}
\includegraphics[scale=0.7]{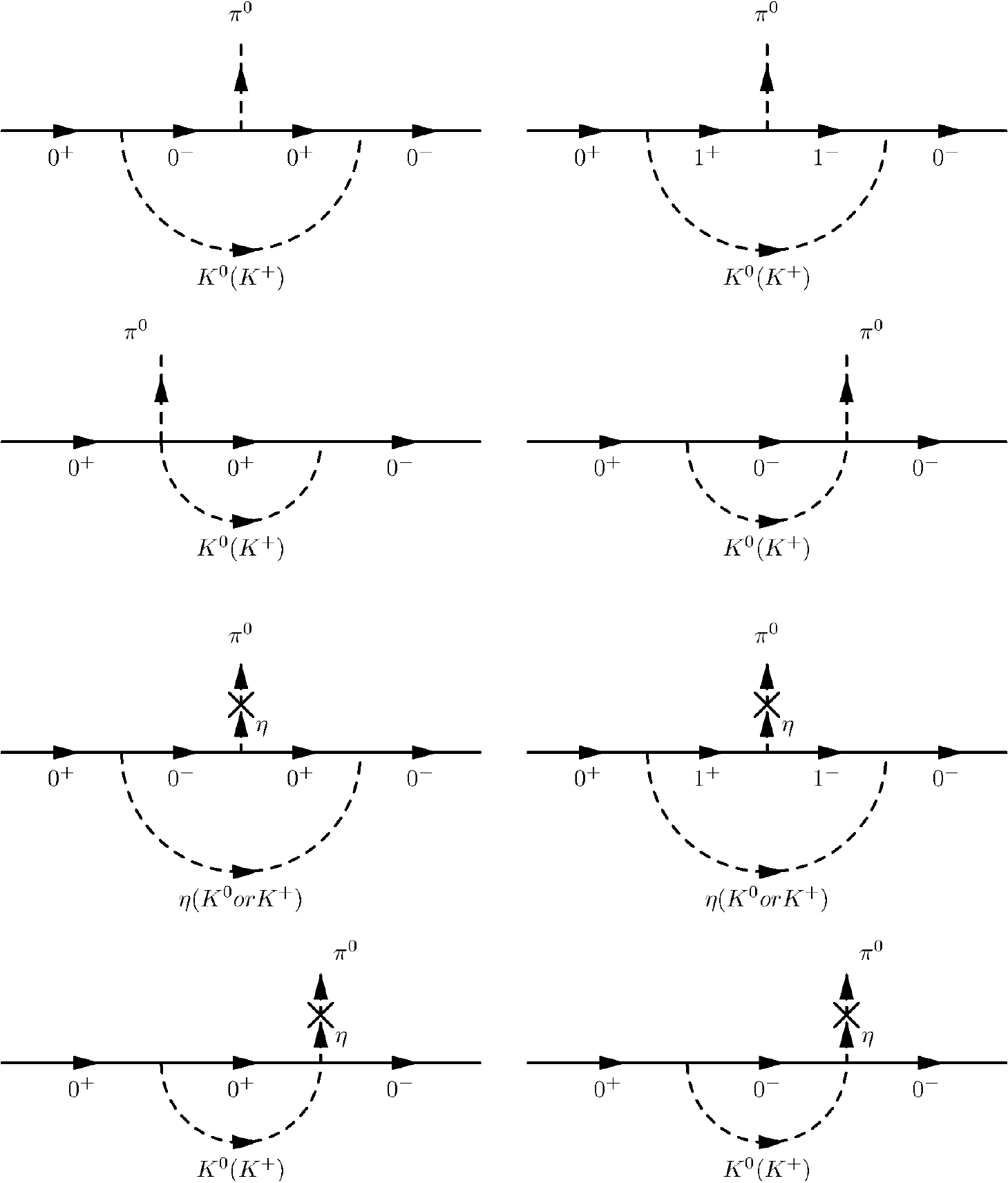}
\caption{Chiral corrections to the $D_{s0}^*(2317)^+ \to D_s^+ \pi^0$ decay mode.}
\label{chi2}
\end{center}
\end{figure}

\begin{figure}
\begin{center}
\includegraphics[scale=0.7]{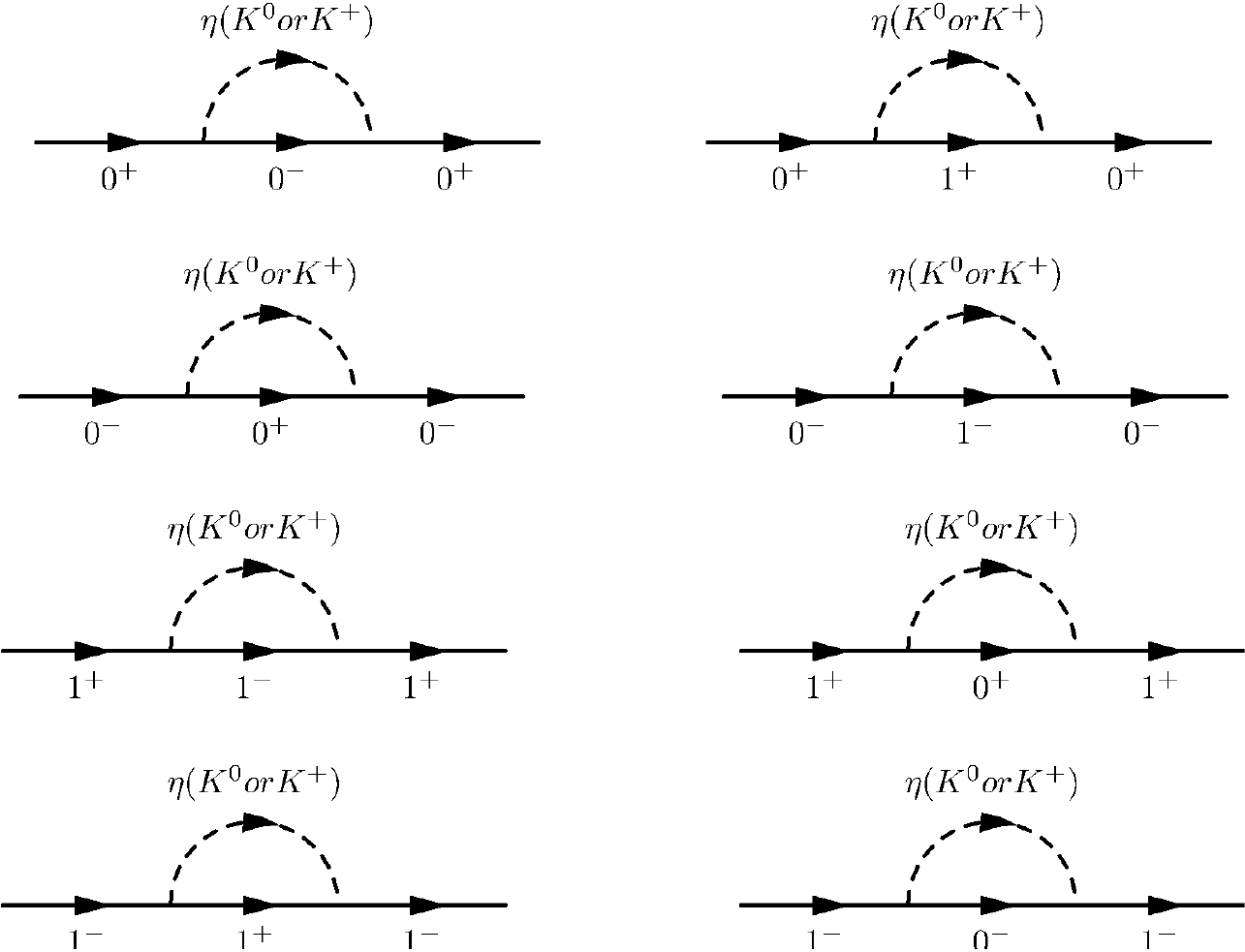}
\caption{Chiral corrections to the $D$ meson wave functions.}
\label{sunrise}
\end{center}
\end{figure}

Next, we calculate chiral loop corrections to the above decay modes. 
By including chiral corrections to the $\eta$-$\pi$ mixing, the  decay width  becomes
\begin{equation}
\Gamma  (D_{s1}(2460)^+ \to D_s^{*+} \pi^0)=\frac{h^2}{2\pi f^2}|k_\pi|E_\pi \delta^2_{mix}\left|\frac{\sqrt{Z_f}\sqrt{Z_i}}{Z_v}\right|^2\,,
\end{equation}
where $Z_f$ and $Z_i$ denote wave function renormalization of the initial and final heavy meson states,
\begin{equation}
Z_{i,f}=1-\frac{1}{2}\frac{\partial \Pi_{i,f} (v \cdot p)}{\partial v \cdot p} \Big |_{on\;mass\; shell}\,,
\end{equation}
while $Z_v$ represents the vertex corrections
\begin{equation}
Z_v=1-\frac{\hat \Gamma(v\cdot p_i,v\cdot p_f,k^2)}{\hat \Gamma_{0}(v\cdot p_i,v\cdot p_f,k^2)} \Big |_{on\;mass\; shell}\,.
\end{equation}
Here $\hat \Gamma$ is the vertex amplitude calculated from the Feynman diagrams presented in Figs. \ref{chi1} and \ref{chi2}, while $\hat \Gamma_0$ is the  vertex amplitude resulting  from the tree-level Feynman diagrams (see Fig. \ref{tree}). Similarly, $\Pi(v\cdot p)$ is the heavy meson self-energy arising from  the Feynman diagrams in Fig. \ref{sunrise}.

The vertex corrections come from the Feynman diagrams presented in Figs. \ref{chi1} and \ref{chi2} and can be summarized into the expression
\begin{equation}
Z_{v}=1-\Big(\delta^\prime_{mix}+\frac{2}{3}{\cal A}_{i}^\prime(m_\eta)-\frac{1}{2}\big({\cal A}_{i}(m_{K^+})+{\cal A}_{i}(m_{K^0})\big)+\frac{1}{\sqrt{2}\delta_{mix}}\big({\cal A}_{i}(m_{K^+})-{\cal A}_{i}(m_{K^0})\big)+{\cal A}_{ct}\Big)\,,
\end{equation}
where $\delta^\prime_{mix}=0.11$  includes corrections to the $\eta-\pi$ mixing angle beyond tree level \cite{f,Stewart:1998ke} and the functions ${\cal A}_i$ are given in Appendix B.

The isospin violating nature of both decays are manifested  either by proportionality of the amplitude  to the mixing parameter $\delta_{mix}$, or by   vanishing of the amplitude in the  case of the isospin limit  $m_{K^0}=m_{K^+}$.

Finite parts of counterterms, are included  in amplitude as  ${\cal A}_{ct}$:
\begin{equation}
{\cal A}_{ct}=\frac{1}{32\pi^2f^2}\left(\left(m_K^2-\frac{m_\pi^2}{2}\right)(\kappa^\prime_1+\kappa^\prime_9)+\left(m_K^2-m_\pi^2+\frac{\sqrt{2}(m^2_{K^+}-m^2_{K^0})}{\delta_{mix}}\right)\kappa^\prime_5+\frac{E_\pi}{2}(\delta_2^\prime+\delta_3^\prime)\right)\,.
\end{equation}
The  values of the finite parts of counterterms, of course, depend on the  renormalization scheme. We use dimensional regularization in the renormalization scheme in which the divergence $2/\epsilon$  contains the constant $-\gamma_E+\ln 4\pi +1$, coming from the loop integrals. This has to be taken into account when discussing the numerical value of the ${\cal A}_{ct}$ term.
The wave function renormalization terms ${\cal Z}_f$ and ${\cal Z}_i$ arise from sunrise diagrams presented in  Fig. \ref{sunrise}
\begin{equation}
Z_j=1-{\cal B}_j(m_{K^+})-{\cal B}_j(m_{K^0})-\frac{2}{3}{\cal B}^\prime_{j}(m_\eta)\,. 
\end{equation}
The functions ${\cal B}_j$ are listed in Appendix B.
\begin{figure}
\includegraphics[scale=0.5]{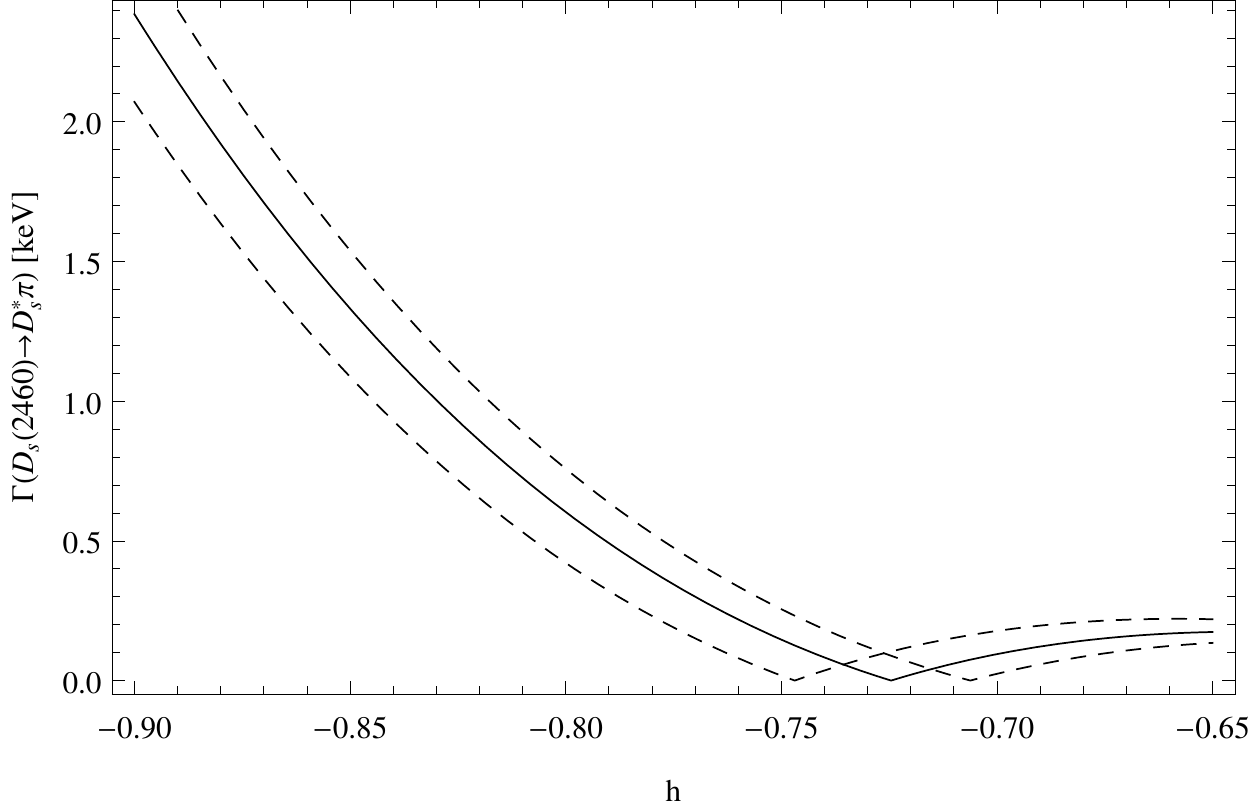}
$\qquad$
\includegraphics[scale=0.5]{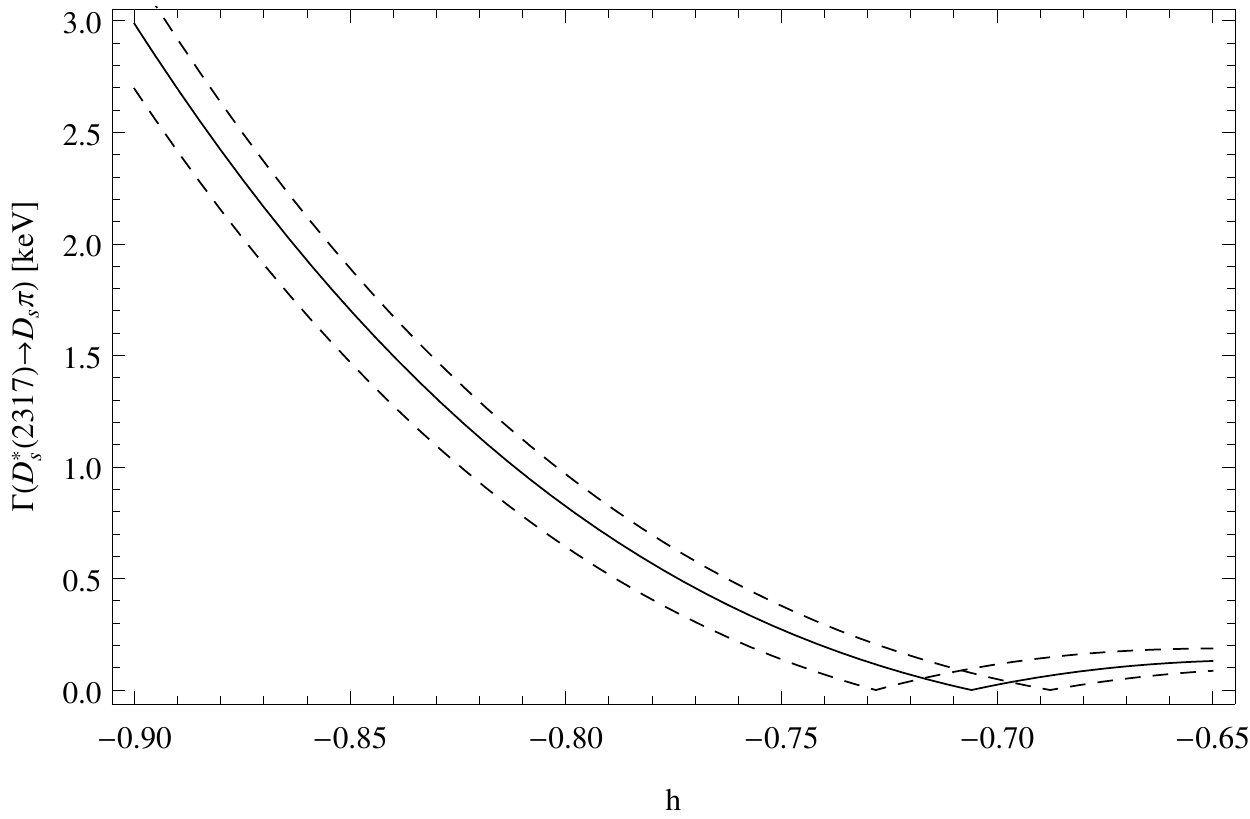} \\
\caption{Dependence of $D_{s1}(2460)^+ \to D_s^{*+} \pi^0$ and $D^*_{s0}(2317)^+ \to D_s^+ \pi^0$ decay widths on input parameters at ${\cal A}_{ct}=0$. Dashed lines present the uncertainty on the decay width due to the uncertainty in $g$ and the mass of the  initial meson.}
\label{napake}
\end{figure}

\begin{figure}
\begin{center}
\includegraphics[scale=0.5]{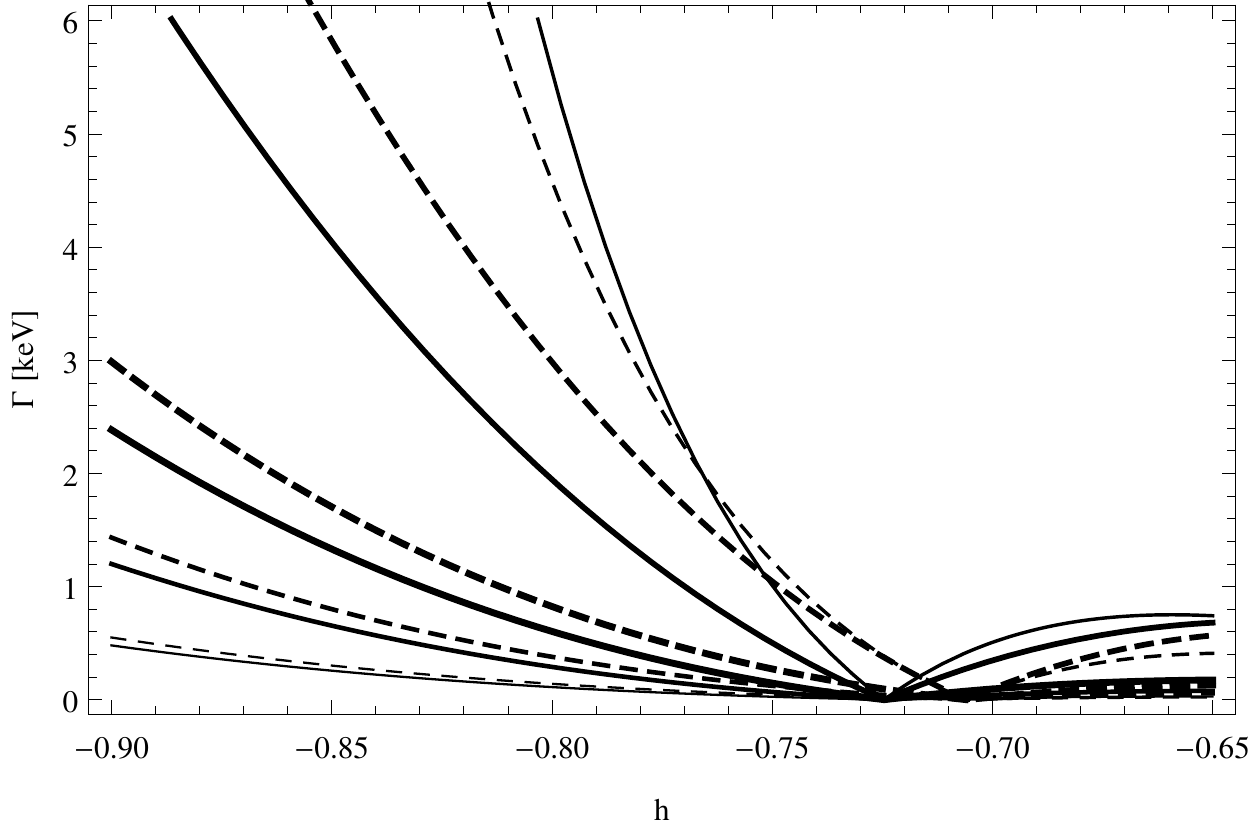}
$\qquad$
\includegraphics[scale=0.5]{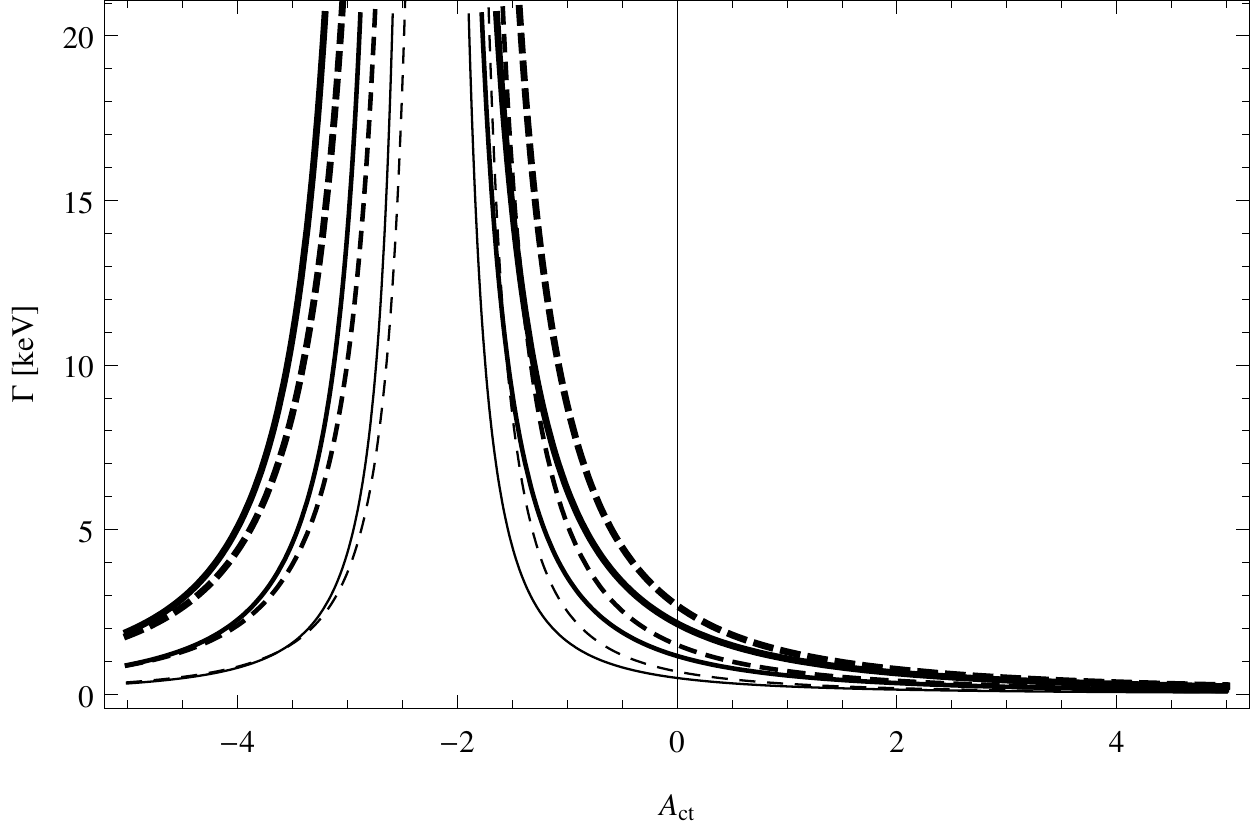}
\caption{Left: Dependence of $D_{s1}(2460)^+ \to D_s^{*+} \pi^0$ and $D^*_{s0}(2317)^+ \to D_s^+ \pi^0$ decay widths on the decay constant $h$ for  five different values of counterterm amplitude. From thickest to thinnest line,  ${\cal A}_{ct}=0,-1,1,-3,3\,$. The $D_{s1}(2460)^+ \to D_s^{*+} \pi^0$ decay width is presented by the solid line, while the $D^*_{s0}(2317)^+ \to D_s^+ \pi^0$ decay width is presented by the dashed line. 
Right: Dependence of $D_{s1}(2460)^+ \to D_s^{*+} \pi^0$ and $D^*_{s0}(2317)^+ \to D_s^+ \pi^0$ decay widths on the value of counterterm amplitude for three different values of coupling constant $h$. From thickest to thinnest line, $h=-0.89,-0.84,-0.79$. The solid line denotes the decay width of $D_{s1}(2460)^+ \to D_s^{*+} \pi^0$, while the dashed line is used  for the decay width of $D^*_{s0}(2317)^+ \to D_s^+ \pi^0$.}
\label{obarazpada}
\end{center}
\end{figure}

The calculated numerical values of the decay widths of $D_{s1}(2460)^+ \to D_s^{*+} \pi^0$ and $D^*_{s0}(2317)^+ \to D_s^+ \pi^0$ are very sensitive on  the value of $h$ and can be  modified by  two orders of magnitude when  $h$ varies form $-0.65$ to $-0.9$. The wave function corrections ${\cal Z}_{i,f}$  are source of that sensitivity. The decay widths are also moderately sensitive on the mass values of final and initial states and vary on coupling constant $g$. 
The dependence on the decay width on  $g$ and the masses of the final and initial mesons are presented in Fig. \ref{napake}, while  the decay widths of $D_{s1}(2460)^+ \to D_s^{*+} \pi^0$ and $D^*_{s0}(2317)^+ \to D_s^+ \pi^0$ are presented in Fig. \ref{obarazpada}, for a range of  values for  counterterms ${\cal A}_{ct}$ .

\begin{figure}
\begin{center}
\includegraphics[scale=0.5]{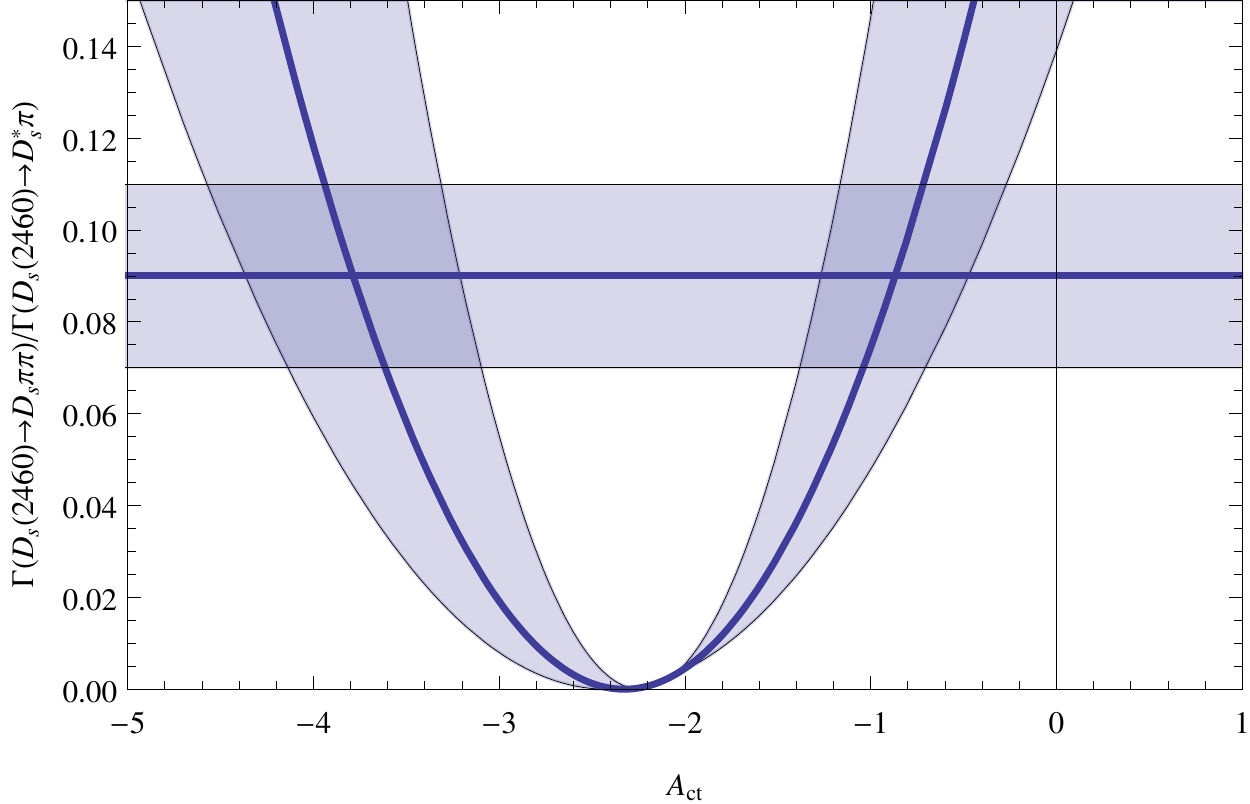}
\caption{Determination of  the allowed regions for  the  counterterm size in the amplitude ${\cal A}_{ct}$. Horizontal  lines present the experimental values of $\frac{BR(D_{s1}(2460)^+\to D_s^{*+}\pi^0)}{BR(D_{s1}(2460)^+\to D_s^{*+}\pi^+\pi^-)}$  with the one sigma error band. The "U" line presents the result of our calculation for the ratio $\frac{BR(D_{s1}(2460)^+\to D_s^{*+}\pi^0)}{BR(D_{s1}(2460)^+\to D_s^{*+}\pi^+\pi^-)}$ with the band region presenting the uncertainty coming  from the variation of $h$ in the range  -0.79 and 0.84.}
\label{kontratermi}
\end{center}
\end{figure}

Since the numerical results are very sensitive on  the value of coupling constant $h$ and due to the unknown final parts of counterterms, we cannot  make a definite  prediction for  the partial decay widths for  both decay modes. Nevertheless, by using experimentally measured ratio  of these rates,
\begin{equation}
\frac{BR(D_{s1}(2460)^+\to D_s^{*+}\pi^0)}{BR(D_{s1}(2460)^+\to D_s^{*+}\pi^+\pi^-)}=0.09\pm 0.02\,,
\end{equation}
we can  shed more light on the value of counterterm amplitude ${\cal A}_{ct}$. 
Our result given  in Fig. \ref{kontratermi},  indicates that  ${\cal A}_{ct}$  can be accommodated within 
the ranges (-4.8;-3.2) and (-1.3;-0.1).
Here, we only considered one $\sigma$ experimental error on the decay widths ratio and the uncertainty in the  theoretical prediction, by varying $h$ between $-0.79$ and $-0.89$.  Within these bounds, we obtain for  the $D_{s1}(2460)^+ \to D_s^{*+} \pi^0$ and $D^*_{s0}(2317)^+ \to D_s^+ \pi^0$ 
decay widths:
\begin{eqnarray}
&&\Gamma(D_{s1}(2460)^+ \to D_s^{*+} \pi^0)= 2.7-3.4\,{\rm keV} \,,\\
&&\Gamma(D^*_{s0}(2317)^+ \to D_s^{+} \pi^0)= 2.4-4.7\,{\rm keV}\,.
\end{eqnarray}

One can notice a slight difference between $\Gamma(D_{s1}(2460)^+ \to D_s^{*+} \pi^0)$ and $\Gamma(D^*_{s0}(2317)^+ \to D_s^{+} \pi^0)$,  not present at the  first order which arises from the $\eta-\pi^0$ mixing. This difference is a result of loop corrections, which were not considered in previous calculations \cite{Colangelo:1997rp}.

\section{Discussion and conclusions}
Within a HM$\chi$PT framework,  we determine  loop contributions to the strong  isospin violating decay amplitudes for  $D_{s1}(2460)^+$ and $D_{s0}^*(2317)^+$. 
We have assumed that both states carry quantum numbers of the  $\bar c s$ states. 
Since  three-body decays of these states are forbidden at tree level, we calculate contributions to the decay amplitude at the loop level, which is consequently finite. 
Contrary to  three-body decay amplitudes described by  the finite loop contributions,  two-body isospin violating decay amplitudes  receive contributions at tree level, induced by the $\eta -\pi^0$ mixing. Therefore, the chiral loop contribution is not finite and in order to regularize it,  one has to introduce counterterms. 
Our estimate of the  size of counterterm in the amplitude ${\cal A}_{ct}$ relies on the result we derive for the  $D_{s1}(2460)^+ \to D_s^+ \pi^+ \pi^-$ decay width. 

Note that within chiral perturbation theory, only the light pseudoscalar mesons are present in the loops. Contributions of the light resonances with the spin $J^P =1^-, \, 0^+, \, 1^+$ are already accounted by the 
chiral loop contributions of  light pseudoscalar mesons \cite{Casalbuoni:1996pg,Ecker:1989yg}. Additional inclusion of  light  resonances  with $J^P =1^-, \, 0^+, \, 1^+$ in HM$\chi$PT 
is not consistent with the original framework. 
Nevertheless, one can roughly estimate contribution of light vector mesons in the loop. For example, the contribution of $K^*$ vector mesons in loops as seen in \cite{Faessler:2007us} would correspond to two loop effects within our framework. Their contribution to the decay rates is therefore suppressed within HM$\chi$PT, in contrast with the $K^*$ loop contribution in the $D^*K$ molecule picture presented in \cite{Faessler:2007us}.

If one assumes as in Ref. \cite{Bardeen:2003kt} that heavy mesons interact with  spin $J^P=0^+$ light resonances as $\sigma$ (or $f_0(500)$ as in \cite{PDG}) at tree level, then $D_{s1}(2460)^+ \to D_s^+ \pi^+ \pi^-$  decay  can proceed through $D_{s1}(2460)^+ \to D_s^+  \sigma \to  D_s^+\pi^+ \pi^-$. However,  their contribution is proportional to the $\epsilon \cdot v$ which is then equal to $0$ in HM$\chi$PT. On the other hand, one might think that since the mass of $\sigma $ is  close to $500\,$ MeV, that state can be important in $D_{s1}(2460)^+ \to D_s^+ \pi^+ \pi^-$. This would be the case if $\sigma$ contained the  $\bar s s$ component. Recent lattice QCD study \cite{Howarth:2015caa} does not support such idea.
In principle, higher order terms in the chiral expansion might contain terms which describe interactions of heavy meson states with $J^P =0^+$ light resonances. At the same time that will mean that some of these $f_0$ states should contain $s \bar s$ contribution. 
Unfortunately, the structure of light positive parity scalar mesons is not known yet  and reliable consideration of this contribution is not possible at present. 

Better understanding of the structure of  $D_{s1}(2460)^+$ and $D_{s0}^*(2317)^+$,  as well as light scalar mesons,  might shed more light on the decay mechanism for  two-  and three-body strong isospin violating decays of  $D_{s1}(2460)^+$ and $D_{s0}^*(2317)^+$.\\

{\bf ACKNOWLEDGMENTS}

The work of S.F. was supported in part by the Slovenian Research Agency. 

\appendix

\section{Loop integrals}

By employing dimensional regularization, in the renormalization scheme with $\delta=\frac{2}{4-D}-\gamma_E+\ln 4\pi +1=0$, we have

$$A_0(m)=\frac{(2\pi\mu)^{4-D}}{i\pi^2}\int\frac{d^Dk}{(k^2-m^2+i\epsilon)}=
m^2\left(\delta-\ln\frac{m^2}{\mu^2}\right)+{\cal O}(D-4)\,,$$
$$B_0(p,m,m)=\frac{(2\pi\mu)^{4-D}}{i\pi^2}\int\frac{d^Dk}{(k^2-m^2+i\epsilon)((k+p)^2-m^2+i\epsilon)}$$$$=\delta-\int_0^1 \ln\frac{x^2p^2-xp^2+m^2}{\mu^2}+{\cal O}(D-4)\,,$$

$$B_{00}(p,m,m)=\frac{1}{2(D-1)}[A_0(m)+(2m^2-p^2/2)B_0(p,m,m)]\,,$$
which in $D \rightarrow 4$ limit gives
$$ B_{00}(p,m,m)=\frac{1}{6}[A_0(m)+(2m^2-p^2/2)B_0(p,m,m)+2m^2-p^2/3]\,,$$
$$B_{00}(m)=B_{00}(\Delta_M v,m,m)\,.$$
Loop integrals with one light meson propagator are
$$\bar B_0(\Delta,m)=\frac{(2\pi\mu)^{4-D}}{i\pi^2}\int\frac{d^Dk}{(k^2-m^2+i\epsilon)(v\cdot k-\Delta+i\epsilon)}=$$$$-2\Delta\left[\delta-\ln\frac{m^2}{\mu^2}-2F\left(\frac{m}{\Delta}\right)+1\right]+{\cal O}(D-4)\,,$$ 
with
$$F(1/x)=\Bigg\{
\begin{matrix} 
\frac{1}{x}\sqrt{x^2-1}\ln(x+\sqrt{x^2-1}+i\epsilon)\,; & |x|>1\,, \\
\frac{-1}{x}\sqrt{1-x^2}\left(\frac{\pi}{2}-\tan^{-1}\left(\frac{x}{\sqrt{1-x^2}}\right)\right)\,; & |x|\leq 1\,, 
\end{matrix}$$

$$\bar B^\mu(\Delta,m)=\frac{(2\pi\mu)^{4-D}}{i\pi^2}\int\frac{k^\mu\, d^Dk}{(k^2-m^2+i\epsilon)(v\cdot k-\Delta+i\epsilon)}=\bar B_1(\Delta,m) v^\mu\,,$$

$$\bar B_1(\Delta,m)=\frac{(2\pi\mu)^{4-D}}{i\pi^2}\int\frac{k \cdot v\, d^Dk }{(k^2-m^2+i\epsilon)(v\cdot k-\Delta+i\epsilon)}=A_0(m)+\Delta \bar B_0(\Delta,m)\,,$$

$$\bar B^{\mu\nu}(\Delta,m)=\frac{(2\pi\mu)^{4-D}}{i\pi^2}\int\frac{k^\mu k^\nu\, d^Dk}{(k^2-m^2+i\epsilon)(v\cdot k-\Delta+i\epsilon)}=\bar B_{00}(\Delta,m) g^{\mu\nu}+\bar B_{11}(\Delta,m) v^\mu v^\nu\,,$$

$$\bar B_{00}(\Delta,m)=\frac{1}{D-1}[(m^2-\Delta^2)\bar B_0(\Delta,m)-\Delta A_0(m)]\,,$$
which in the $D\rightarrow 4$ gives
$$\bar B_{00}(\Delta,m)=\frac{1}{3}[(m^2-\Delta^2)\bar B_0(\Delta,m)-\Delta A_0(m)+2\Delta/3(3m^2-2\Delta^2)]\,,$$

$$\bar B_{11}(\Delta,m)=\frac{1}{D-1}[(D\Delta^2-m^2)\bar B_0(\Delta,m)+D\Delta A_0(m)]\,,$$
which in $D\rightarrow 4$ gives
$$\bar B_{11}(\Delta,m)=\frac{1}{3}[(4\Delta^2-m^2)\bar B_0(\Delta,m)+4\Delta A_0(m)-2\Delta/3(3m^2-2\Delta^2)]\,,$$
$$\bar B_{2}(\Delta,m)=\bar B_{00}(\Delta,m)+\bar B_{11}(\Delta,m)\,,$$

$$\bar B_0^\prime(\Delta_1,\Delta_2,m)=\frac{(2\pi\mu)^{4-D}}{i\pi^2}\int\frac{d^Dk}{(k^2-m^2)(v\cdot k-\Delta_1)(v\cdot k-\Delta_2)}=\frac{1}{\Delta_1-\Delta_2}[\bar B_0(\Delta_1,m)-\bar B_0(\Delta_2,m)]\,,$$

$$\bar B^{\mu\prime}(\Delta_1,\Delta_2,m)=\frac{(2\pi\mu)^{4-D}}{i\pi^2}\int\frac{k^\mu\, d^Dk}{(k^2-m^2)(v\cdot k-\Delta_1)(v\cdot k-\Delta_2)}=\bar B_1^\prime(\Delta_1,\Delta_2,m) v^\mu\,,$$

$$\bar B_1^\prime(\Delta_1,\Delta_2,m)=\frac{(2\pi\mu)^{4-D}}{i\pi^2}\int\frac{k\cdot v\, d^Dk}{(k^2-m^2)(v\cdot k-\Delta_1)(v\cdot k-\Delta_2)}=\bar B_0(\Delta_2,m)+\Delta_1 \bar B_0^\prime(\Delta_1,\Delta_2,m)\,,$$

$$\bar B_2^\prime(\Delta_1,\Delta_2,m)=\frac{(2\pi\mu)^{4-D}}{i\pi^2}\int\frac{(k\cdot v)^2\, d^Dk}{(k^2-m^2)(v\cdot k-\Delta_1)(v\cdot k-\Delta_2)}=$$$$A_0(m)+(\Delta_1+\Delta_2)\bar B_0(\Delta_2,m)+\Delta_1^2 \bar B_0^\prime(\Delta_1,\Delta_2,m)\,,$$

$$\bar B^{\mu\nu\prime}(\Delta_1,\Delta_2,m)=\frac{(2\pi\mu)^{4-D}}{i\pi^2}\int\frac{k^\mu k^\nu\, d^Dk}{(k^2-m^2)(v\cdot k-\Delta_1)(v\cdot k-\Delta_2)}=$$$$\bar B_{00}^\prime(\Delta_1,\Delta_2,m) g^{\mu\nu}+\bar B_{11}^\prime(\Delta_1,\Delta_2,m) v^\mu v^\nu\,,$$

$$\bar B_{00}^\prime(\Delta_1,\Delta_2,m)=\frac{1}{D-1}[m^2\bar B_{0}^\prime(\Delta_1,\Delta_2,m)-\Delta_1\bar B_{1}^\prime(\Delta_1,\Delta_2,m)-\bar B_{1}(\Delta_2,m)]\,,$$
which in $D\rightarrow 4$ gives
$$\frac{1}{3}[m^2\bar B_{0}^\prime(\Delta_1,\Delta_2,m)-\Delta_1\bar B_{1}^\prime(\Delta_1,\Delta_2,m)-\bar B_{1}(\Delta_2,m)+2/3(3m^ 2-2(\Delta_1^2+\Delta_2^2+\Delta_1\Delta_2))]\,,$$

$$\bar B_{11}^\prime(\Delta_1,\Delta_2,m)=\frac{1}{D-1}[-m^2\bar B_{0}^\prime(\Delta_1,\Delta_2,m)+D\Delta_1 \bar B_{1}^\prime(\Delta_1,\Delta_2,m)+D\bar B_{1}(\Delta_2,m)]\,,$$
which in $D\rightarrow 4$ gives
$$\frac{1}{3}[-m^2\bar B_{0}^\prime(\Delta_1,\Delta_2,m)+4\Delta_1 \bar B_{1}^\prime(\Delta_1,\Delta_2,m)+4\bar B_{1}(\Delta_2,m)-2/3(3m^ 2-2(\Delta_1^2+\Delta_2^2+\Delta_1\Delta_2))]\,,$$

Loop integrals with two light meson propagators are
$$\bar C^\mu(p,\Delta,m_1,m_2)=\frac{(2\pi\mu)^{4-D}}{i\pi^2}\int\frac{k^\mu\, d^Dk}{(k^2-m_1^2+i\epsilon)((k-p)^2-m_2^2+i\epsilon)(v\cdot k-\Delta+i\epsilon)}=$$$$\bar C_1(p,\Delta,m_1,m_2) v^\mu\,,$$

$$\bar C_1(p,\Delta,m_1,m_2)=\frac{(2\pi\mu)^{4-D}}{i\pi^2}\int\frac{k\cdot v\, d^Dk}{(k^2-m_1^2+i\epsilon)((k-p)^2-m_2^2+i\epsilon)(v\cdot k-\Delta+i\epsilon)}=$$$$B_0(p,m_1,m_2)+\Delta\bar C_0(p,\Delta,m_1,m_2)\,,$$

$$\bar C^{\mu\nu}(p,\Delta,m_1,m_2)=\frac{(2\pi\mu)^{4-D}}{i\pi^2}\int\frac{k^\mu k^\nu\, d^Dk}{(k^2-m_1^2+i\epsilon)((k-p)^2-m_2^2+i\epsilon)(v\cdot k-\Delta+i\epsilon)}=$$$$\bar C_{00}(p,\Delta,m_1,m_2)g^{\mu\nu}+ \bar C_{11}(p,\Delta,m_1,m_2)v^\mu v^\nu\,,$$

$$\bar C_{00}(\Delta,m)=\bar C_{00}(-\Delta_M v,\Delta,m,m)=\frac{1}{D-1}[\bar B_0(-\Delta_M+\Delta,m)-(\Delta_M/2+\Delta) B_0(\Delta_m v,m,m)+$$$$(m^2-\Delta^2)\bar C_0(\Delta_M v,\Delta,m,m)]\,,$$
which in the $D\rightarrow 4$ gives
$$\bar C_{00}(\Delta,m)=\bar C_{00}(-\Delta_M v,\Delta,m,m)=\frac{1}{3}[\bar B_0(-\Delta_M+\Delta,m)-(\Delta_M/2+\Delta) B_0(\Delta_m v,m,m)+$$$$(m^2-\Delta^2)\bar C_0(\Delta_M v,\Delta,m,m)-2/3(3/2\Delta_M-\Delta)]\,,$$

$$\bar C_{11}(\Delta,m)=\bar C_{11}(-\Delta_M v,\Delta,m,m)=\frac{1}{D-1}[-\bar B_0(-\Delta_M+\Delta,m)+D(\Delta_M/2+\Delta) B_0(\Delta_m v,m,m)-$$$$(m^2-D\Delta^2)\bar C_0(\Delta_M v,\Delta,m,m)]\,,$$
which in $D\rightarrow 4$ gives
$$\bar C_{11}(\Delta,m)=\bar C_{11}(-\Delta_M v,\Delta,m,m)=\frac{1}{3}[-\bar B_0(-\Delta_M+\Delta,m)+4(\Delta_M/2+\Delta) B_0(\Delta_m v,m,m)-$$$$(m^2-4\Delta^2)\bar C_0(\Delta_M v,\Delta,m,m)+2/3(3/2\Delta_M-\Delta)]\,.$$
The calculation of the integral,
$$\bar C_0(p,\Delta,m_1,m_2)=\frac{(2\pi\mu)^{4-D}}{i\pi^2}\int\frac{d^Dk}{(k^2-m_1^2+i\epsilon)((k-p)^2-m_2^2+i\epsilon)(v\cdot k-\Delta+i\epsilon)}\,$$
is done in \cite{Zupan:2002je}.
For some calculations, we used the program FeynCalc \cite{feyncalc}.

\section{Loop functions}

Loop functions entering Eq. (20) are listed here: 
$$
{\cal A}_1(m_i)=\frac{1}{16\pi^2f^2}\left((\bar B_{00}(\Delta_{P^*},m_i)+\bar B_{11}(\Delta_{P^*},m_i)-\Delta_M\bar B_{1}(\Delta_{P^*},m_i)\right.
$$$$\left.-\bar B_{00}(\Delta_{D^*}+\Delta_M,m_i)-\bar B_{11}(\Delta_{D^*}+\Delta_M,m_i)-\Delta_M\bar B_{1}(\Delta_{D^*}+\Delta_M,m_i))/2\right.
$$$$\left.-h^2\left(\bar B^\prime_{00}(\Delta_{P^*},\Delta_{D^*}+\Delta_M,m_i)+\bar B^\prime_{11}(\Delta_{P^*},\Delta_{D^*}+\Delta_M,m_i)\right)\right.+
$$
\begin{equation}
+\left. g\tilde g\left(\bar B^\prime_{00}(\Delta_{P},\Delta_{D}+\Delta_M,m_i)+2\bar B^\prime_{00}(\Delta_{D^*},\Delta_{P^*}+\Delta_M,m_i)\right)\right)\,,
\end{equation}
$$
{\cal A}^\prime_1(m_i)=\frac{1}{16\pi^2f^2}\left(
-h^2\left(\bar B^\prime_{00}(\Delta_{P^*},\Delta_{D^*}+\Delta_M,m_i)+\bar B^\prime_{11}(\Delta_{P^*},\Delta_{D^*}+\Delta_M,m_i)\right)\right.+
$$
\begin{equation}
+\left. g\tilde g\left(\bar B^\prime_{00}(\Delta_{P},\Delta_{D}+\Delta_M,m_i)+2\bar B^\prime_{00}(\Delta_{D^*},\Delta_{P^*}+\Delta_M,m_i)\right)\right)\,,
\end{equation}
for the $D_{s1}(2460)^+ \to D_s^{*+} \pi^0$ decay mode and
$$
{\cal A}_2(m_i)=\frac{1}{16\pi^2f^2}\left((\bar B_{00}(\Delta_{P},m_i)+\bar B_{11}(\Delta_{P},m_i)-\Delta_M\bar B_{1}(\Delta_{P},m_i)\right.
$$$$\left.-\bar B_{00}(\Delta_{D}+\Delta_M,m_i)-\bar B_{11}(\Delta_{D}+\Delta_M,m_i)-\Delta_M\bar B_{1}(\Delta_{D}+\Delta_M,m_i))/2\right.
$$$$\left.-h^2\left(\bar B^\prime_{00}(\Delta_{P},\Delta_{D}+\Delta_M,m_i)+\bar B^\prime_{11}(\Delta_{P},\Delta_{D}+\Delta_M,m_i)\right)\right.+$$
\begin{equation}
+\left. 3g\tilde g\bar B^\prime_{00}(\Delta_{D^*},\Delta_{P^*}+\Delta_M,m_i)\right)\,,
\end{equation}

$$
{\cal A}^\prime_2(m_i)=\frac{1}{16\pi^2f^2}\left(
-h^2\left(\bar B^\prime_{00}(\Delta_{P},\Delta_{D}+\Delta_M,m_i)+\bar B^\prime_{11}(\Delta_{P},\Delta_{D}+\Delta_M,m_i)\right)\right.+$$
\begin{equation}
+\left. 3g\tilde g\bar B^\prime_{00}(\Delta_{D^*},\Delta_{P^*}+\Delta_M,m_i)\right)\,,
\end{equation}
for the $D^*_{s0}(2317)^+ \to D_s^+ \pi^0$  decay mode.

\begin{table}
\begin{tabular}{|c|ccccc|}
\hline
In the case of & $\Delta_P$ [GeV] & $\Delta_{P^*}$ [GeV] & $\Delta_D$ [GeV] & $\Delta_{D^*}$ [GeV]   & $\Delta_M$ [GeV]\\ \hline
${\cal A}_1$ & -0.59 &  -0.45 & -0.06 & -0.04 & -0.35  \\
${\cal A}_2$ & -0.48 &  -0.31 &  0.09 & 0.1 &  -0.33 \\ 
${\cal A}^\prime_1$ & -0.47 &  -0.35 & -0.14 & 0 & -0.35  \\
${\cal A}^\prime_2$ & -0.33 &  -0.21 & 0 & -0.14 & -0.33  \\ \hline
\end{tabular}
\caption{Mass differences}
\label{mase}
\end{table}

The functions ${\cal B}_j$ which are present in Eq.(22) are:
\begin{equation}
{\cal B}_{D_s^+}(m_i)=\frac{1}{16\pi^2f^2}\Big(3\tilde g^2\bar B^\prime_{00}(m_{D^*}-m_{D_s^+},m_i)-
h^2\big(\bar B^\prime_{00}(m_{P}-m_{D_s^+},m_i)+\bar B^\prime_{11}(m_{P}-m_{D_s^+},m_i)\big)\Big)\,,
\end{equation}
$$
{\cal B}_{D_s^{*+}}(m_i)=\frac{1}{16\pi^2f^2}\Big(3\tilde g^2\big(-\bar B^\prime_{00}(m_{D}-m_{D_s^{*+}},m_i)+2
\bar B^\prime_{00}(m_{D^*}-m_{D_s^{*+}},m_i)\big)$$
\begin{equation}
-h^2\big(\bar B^\prime_{00}(m_{P^*}-m_{D_s^{*+}},m_i)+\bar B^\prime_{11}(m_{P^*}-m_{D_s^{*+}},m_i)\big)\Big)\,,
\end{equation}
\begin{equation}
{\cal B}_{P_s}(m_i)=\frac{1}{16\pi^2f^2}\Big(3g^2\bar B^\prime_{00}(m_{P^*}-m_{P_s},m_i)
-h^2\big(\bar B^\prime_{00}(m_{D}-m_{P_s},m_i)+\bar B^\prime_{11}(m_{D}-m_{P_s},m_i)\big)\Big)\,,
\end{equation}
$${\cal B}_{P^*_s}(m_i)=\frac{1}{16\pi^2f^2}\Big(3g^2\big(-\bar B^\prime_{00}(m_{P}-m_{P^*_s},m_i)+
2\bar B^\prime_{00}(m_{P^*}-m_{P^*_s},m_i)\big)+$$
\begin{equation}
-h^2\big(\bar B^\prime_{00}(m_{D^*}-m_{P^*_s},m_i)+\bar B^\prime_{11}(m_{D^*}-m_{P^*_s},m_i)\big)\Big)\,.
\end{equation}
The expressions for ${\cal B}^\prime_j(m_i)$ can be obtained from the above expressions of ${\cal B}_j(m_i)$ by substituting masses of $D$ mesons by the masses of $D_s$ mesons.

Here, $\bar B_{1}$,  $\bar B_{00}$ and $\bar B_{11}$, $\bar B^\prime_{00}$ and $\bar B^\prime_{11}$ are loop integrals defined in Appendix A.
Mass differences $\Delta_D$, $\Delta_P$, $\Delta_{D^*}$ $\Delta_{P^*}$ are defined as a mass differences between the appropriate state and the initial state while $\Delta_M$ is the mass difference between final and initial state.  Therefore, mass differences entering different amplitudes are not the same. The values are given in Table \ref{mase}.

\end{document}